\documentclass[%
reprint,
superscriptaddress,
showpacs,preprintnumbers,
 amsmath,amssymb,
 aps,
]{revtex4-1}
\usepackage{color}
\usepackage{lineno}
\usepackage{graphicx}
\usepackage{dcolumn}
\usepackage{bm}
\usepackage{braket}

\begin{document}

\title{Improving Resolution of Solid State NMR in Dense Molecular Hydrogen}

\author{Thomas Meier}
   \email{thomas.meier@uni-bayreuth.de}
   \affiliation{Bavarian Geoinstitute, University of Bayreuth, D-95447 Bayreuth, Germany}

\author{Saiana Khandarkhaeva}
  \affiliation{Bavarian Geoinstitute, University of Bayreuth, D-95447 Bayreuth, Germany}

\author{Jeroen Jacobs}
  \affiliation{European Synchrotron Radiation Facility (ESRF), F-38043 Grenoble Cedex, France}                         
  
\author{Natalia Dubrovinskaia}
  \affiliation{Material Physics and Technology at Extreme Conditions, Laboratory of Crystallography, University of Bayreuth, D-95447 Bayreuth, Germany}                         

\author{Leonid Dubrovinksy}
  \affiliation{Bavarian Geoinstitute, University of Bayreuth, D-95447 Bayreuth, Germany}  

\date{\today}

\begin{abstract}
Recent advancements in radio-frequency resonator designs have led to the implementation of nuclear magnetic resonance in diamond anvil cells (DACs) at pressures well above 100 GPa. However, a relatively low resolution and the absence of decoupling sequences complicate the analysis of the results of solid state NMR in DACs. Here, we present the first application of homo-nuclear Lee-Goldburg (LG) decoupling at extreme conditions on high density molecular hydrogen. Lenz lens based two-dimensional resonator structures were found to generate a homogeneous B$_1$ field across sample cavities as small as 12 picolitres, a prerequisite for optimal decoupling. At ideal LG conditions, the broad $^1$H resonance of molecular ortho-hydrogen was narrowed 1600-fold, resulting in line-widths of 3.1 ppm. 
\end{abstract}

\maketitle
 
After the first prediction of a metallic phase under high compression by Wigner and Huntington\cite{Wigner1935}, and an estimation of the superconducting properties at room temperature by Ashcroft\cite{Ashcroft1968}, H$_2$ became an archetypical test system in modern high pressure science\cite{Mao1994}.  In 2017, experimental\cite{Dias2016a} observations (although heavily disputed\cite{Eremets2017, Liu2017, Goncharov2017} suggested the transformation of hydrogen into a metallic state at around 500 GPa. Recent IR spectroscopic results\cite{Loubeyre2019} indicate a band gap closure at a pressure of 425 GPa, which may be interpreted as the onset of metallisation.\\
Molecular hydrogen remains an elusive “quantum solid”\cite{VanKranendonk1983} in high pressure research. Owing to the lack of core electrons, hydrogen nuclei are practically invisible to X-ray diffraction methods\cite{Stojilovic2012}. The equation of state of solid hydrogen was, however, determined up to 1 Mbar following three sets of diffraction peaks (100, 002 and 101)\cite{Loubeyre1996}. Other structural probes, such as neutron diffraction, cannotbe applied at the necessary pressures in excess of 100 GPa\cite{Boehler2013}, and complementary spectroscopic methods such as IR or Raman provide only limited information on the atomic or molecular arrangements of dense hydrogen\cite{Mazin1997}. \\
Nuclear Magnetic Resonance spectroscopy (NMR) is an alternative to the techniques mentioned above, and is widely regarded as one of the most versatile spectroscopic methods, probing local magnetic fields of single nuclei or whole molecular building blocks within inorganic compounds. A reason for the wide applicability of NMR lies in the development of Average-Hamiltonian-Theory\cite{Brinkmann2016} and subsequent decoupling pulse sequences\cite{Mehring1972, Vinogradov2001} as well as mechanical rotation methods\cite{Lowe1959, Andrew1958}, leading to resonance line narrowing from formerly several hundred or thousands of ppm to often sub-ppm resolution\cite{Hennel2005}. While such high resolution techniques find widespread use in the fields of biochemistry, chemistry, or soft-matter physics, an application within diamond anvil cell research has never been pursued. \\
Recent developments of high pressure NMR techniques, using three-dimensional microcoils\cite{Meier2017b} or two-dimensional Lenz-lens based resonators\cite{Meier2017, Meier2018, Meier2018a}, enabled the detection of faint nuclear induction signals at pressures well into the megabar (1 Mbar = 100 GPa) regime with record pressures of  up to 202 GPa, as reported recently\cite{Meier2019a}. Most of these high pressure NMR experiments have been focused on $^{1}$H nuclei\cite{Meier2018b}, owing to the large gyromagnetic ratio of protons, and subsequently high signal-to-noise ratios\cite{Levitt2000}. At such pressures, a plethora of exotic physical phenomena can be observed, such as the elusive hydrogen-bond symmetrisation from ice VII to ice X governed by quantum-mechanical tunneling of protons within the H-bonds\cite{Meier2018}, or pressure induced hydrogen-hydrogen interactions in metal hydrides as a precursor of high temperature superconductivity\cite{Meier2019a}. \\
Monserrat et al.\cite{Monserrat2019} pointed out that  NMR might be the only method available in diamond anvil cell based research to directly probe the atomic arrangement of hydrogen atoms at high densities. Using ab-initio calculations, theoretical $^{1}$H-NMR spectra at megabar pressures have been calculated, showing a surprisingly large dispersion of chemical shielding values, yielding the possibility to distinguish between theoretical high pressure phases of dense hydrogen.However, an attempt to identify crystallographic phases of H$_2$ seemed unreasonable, as $^{1}$H-NMR linewidths in excess of $10^{3}$ ppm for ortho-hydrogen have been reported under cryogenic temperatures\cite{Hardy1973, Berlinsky1973, Washburn1983, Pedroni1976}. Under decreasing H-H distances, this effect would certainly be more pronounced in high density phases of H$_2$. \\
Here we present the first application of homo-nuclear Lee-Goldburg decoupling\cite{Lee1965} of molecular ortho-hydrogen in diamond anvil cells at pressures between 20 and 64 GPa.\\
\begin{figure*}[htb]
  \includegraphics[width=1.75\columnwidth]{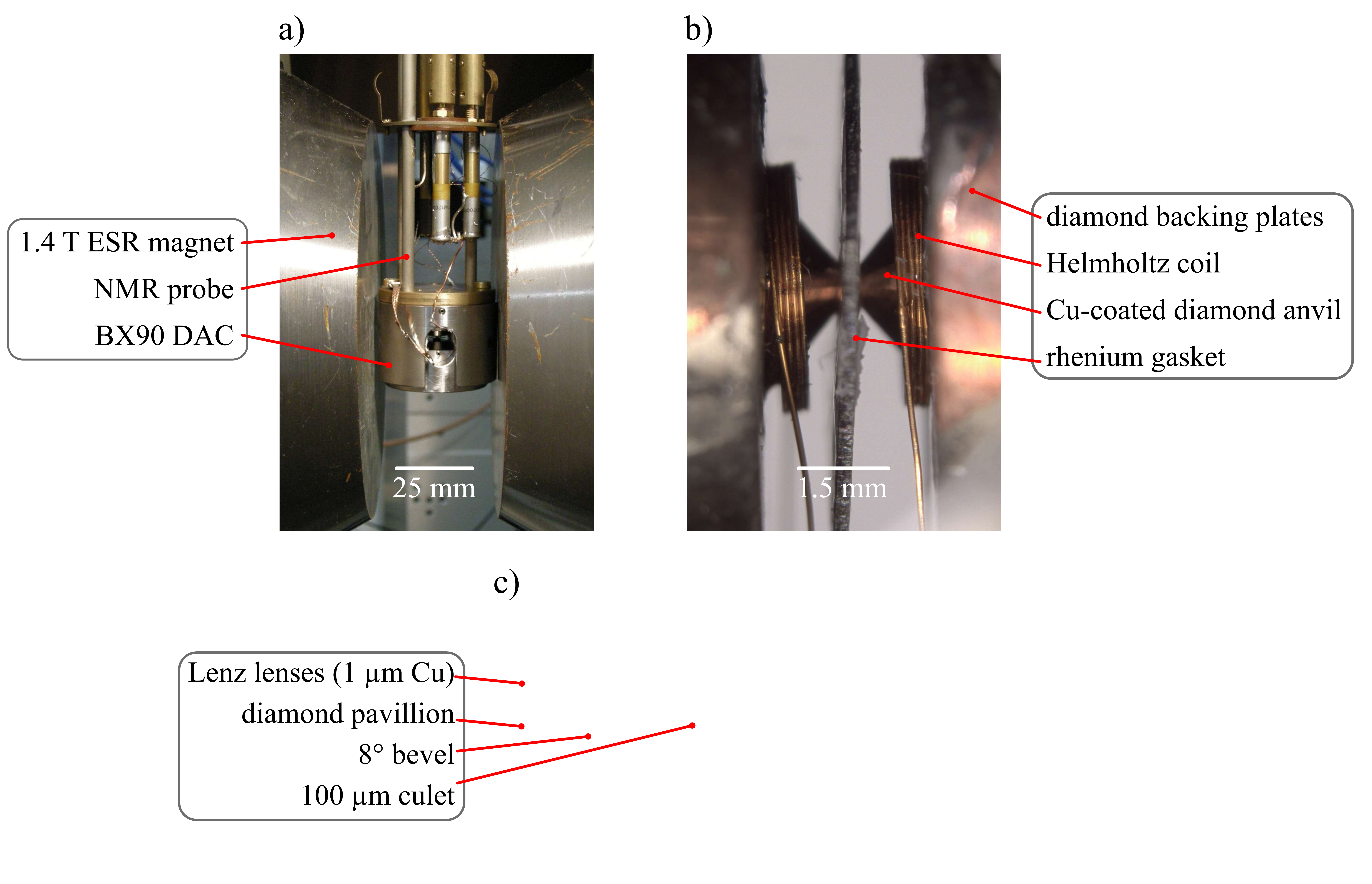}%
 \caption{Experimental high pressure NMR set-up. a) NMR probe with mounted DAC placed in the centre of a 1.4 T ESR magnet. b) view inside a DAC for NMR experiments. The Helmholtz coil has been made from 100 $\mu$m PTFE insulated copper wire. A 1 $\mu$m copper layer was deposited on the diamonds’ surface using physical vapor deposition. The rhenium gasket was electrically insulated using chemical vapor deposition of a 500 nm thick layer of  Al$_2$O$_3$. c) SEM photograph of a 100 $\mu$m culet diamond with three Lenz lenses on the diamond’s pavilion, the beveled region as well as the culet.
 \label{FIG1}}
 \end{figure*} 
Two DACs with 100 $\mu$m and 250 $\mu$m culeted diamonds have been prepared. Rhenium gaskets were pre-indented to about 15 and 25 $\mu$m thickness and the central hole has been laser drilled into the centre of the flat indented area, forming the sample cavity. Subsequently, the diamonds have been coated with a layer of 1 $\mu$m copper, and the rhenium gaskets with a 500 nm layer of electrically insulating Al$_2$O$_3$. Two dimensional Lenz lens NMR resonators have been cut out of the copper layers using focused ion beam milling.  \\
All excitation coils were manufactured from 100 $\mu$m PTFE insulated copper wire, each consisting of 5 turns, and fixed around the diamond anvils. After loading, both coils were combined to form a Helmholtz-like excitation coil, providing a homogeneous inductive coupling into the respective Lenz lenses. Figure 1 shows three photographs of the set-up used: BX90\cite{Kantor2012} pressure cell mounted on a home-built NMR probe in the region of highest homogeneity of an ESR magnet (a); photograph of the inside of the DAC with both copper coated diamonds, the sealing rhenium gasket, as well as the Helmholtz excitation coil (b) and an SEM photograph of a prepared triple stage Lenz lens on the surface of a 100 $\mu$m culeted diamond.  \\
Pressure cells were loaded with molecular hydrogen at the European Synchrotron Radiation Facility. Pressure was increased under cryogenic conditions by submerging the cells into liquid nitrogen, and Raman spectra were recorded at room temperature after every pressure increase, see figure 2. The pressure induced shift of the recorded vibron spectra at about 4250 cm$^{-1}$ are in accordance with previous studies\cite{Sharma1980, Hemley1988}. \\
\begin{figure}[htb]
  \includegraphics[width=0.6\columnwidth]{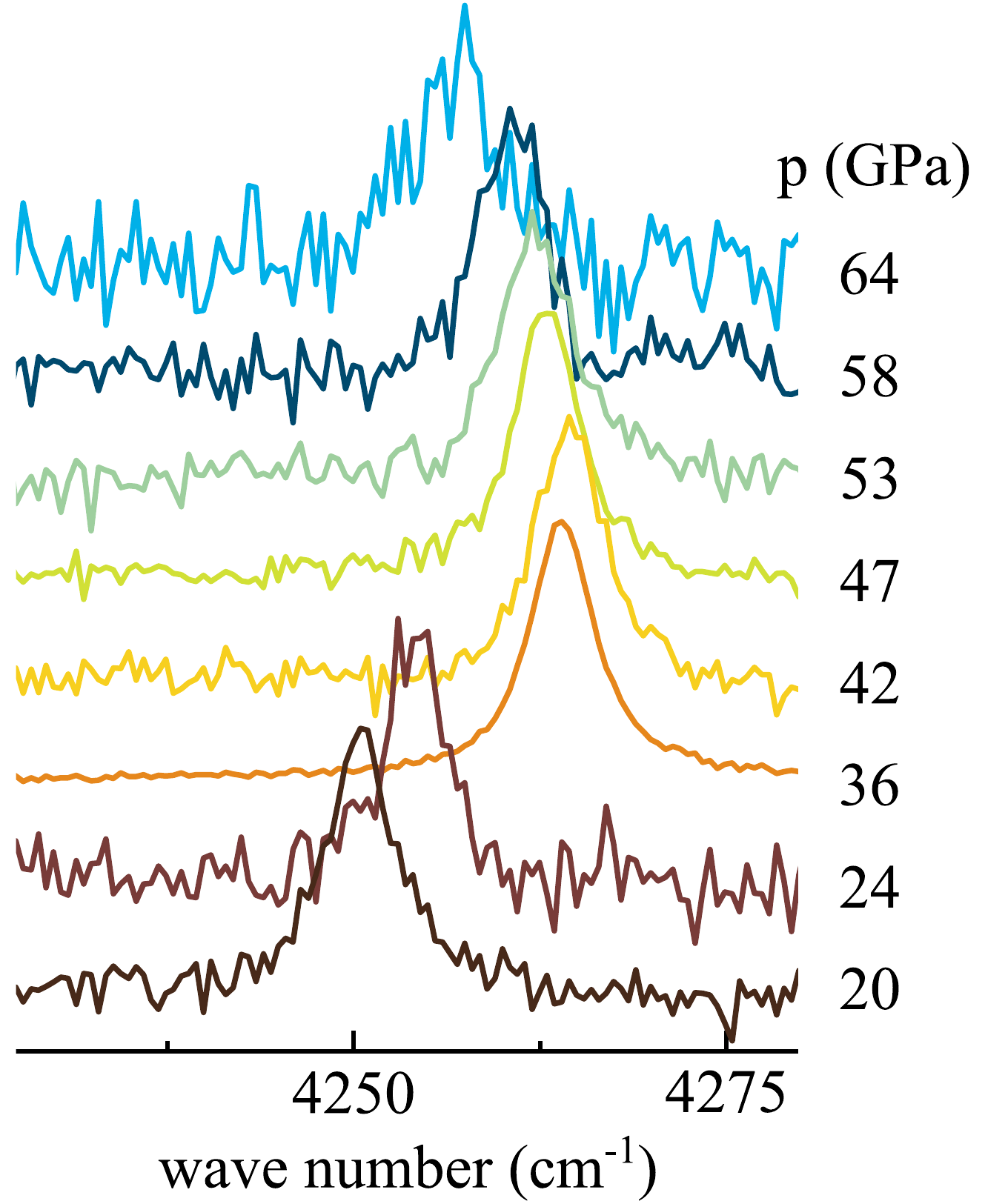}%
 \caption{Raman spectra of molecular H$_2$ under increasing pressure.
 \label{FIG2}}
 \end{figure} 
All NMR experiments have been conducted at an external magnetic field of about 1063 mT, corresponding to a proton resonance frequency of 45.26 MHz. Chemical shift referencing has been performed, using a water sample as a reference standard in a BX90 pressure cell prepared in the same way as the hydrogen loaded cells. \\
Performing $^1$H-NMR nutation experiments at an average pulse power of 8 W has shown optimal 90$^{\circ}$ pulse lengths of about 0.4 to 0.5 $\mu$s, corresponding to RF fields of $B_1=\pi/(2\gamma_n t_{\pi/2}) =$ 11 to 15 mT. Supplementary figure 1 shows numerical simulations, using the $FEMM$ software package, of the B$_1$ field distribution across a 40 $\mu$m x 15 $\mu$m sample cavity used when 100 $\mu$m culeted anvils are used. An average value of  $\braket{B_1} = 17 mT$ (green iso-lines in supplementary figure 1) was found, in excellent agreement with the nutation experiments.  Considering a standard deviation of about $\sigma$ = 3 mT (red isolines in supplementary figure 1), it was found that approximately 80 \% of the sample volume lies within $\braket{B_1} \pm \sigma$. Thus, an effective volume $V_{eff}$, which is subject to a homogeneous $B_1$ field of $V_{eff}=0.8 \cdot V_0$, has been defined.\\
\begin{figure}[htb]
  \includegraphics[width=1\columnwidth]{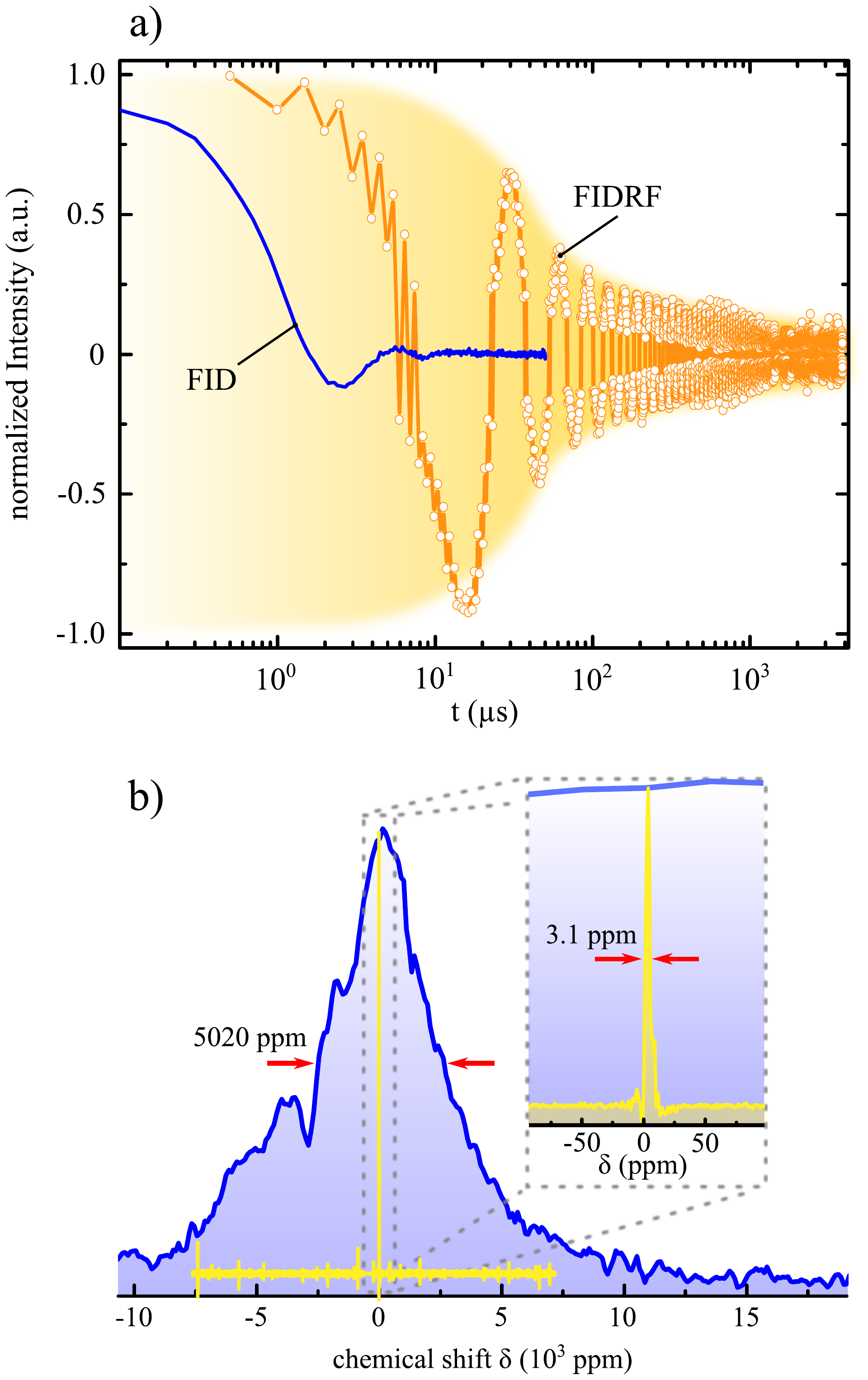}%
 \caption{a) time domain data of the standard FID (blue) and of the FIDRF (orange) after optimisation of LG conditions. The yellow envelope is guide to the eye. B) corresponding $^1$H-NMR spectra under standard “static” conditions (blue) and under LG conditions (yellow), the inset shows a zoomed-in region of the narrowed NMR signal.
 \label{FIG3}}
 \end{figure} 
According to the equation of state of molecular hydrogen in phase I, the molar volume at 20 GPa is 5.25 cm$^{3}/mol$\cite{Loubeyre1996}. Therefore, an upper limit of a total amount of 3 nmol of hydrogen loaded in the sample cavities could be estimated, corresponding to 1.7 $\cdot$ 10$^{15}$ H$_2$ molecules. Considering a normalised time domain signal-to-noise ratio of 8.8 and a limiting bandwidth of 1 MHz, a limit-of-detection\cite{Meier2017} of 2$\cdot$10$^{11}$ spins/Hz$^{1/2}$ has been found. \\
Performing standard single pulse experiments, an intense and rapidly decaying $^{1}$H signal was detected, at the respective Larmor frequency of protons at 1063 mT, figure 3a. Fourier transform results in a dipolar broadened resonance spectrum of about 5000 ppm line width, figure 3b. Using inversion recovery experiments, spin lattice relaxation times between 15 to 21 ms have been found, in accordance with other NMR studies of Phase I of H$_2$ at cryogenic temperatures\cite{Hardy1973, Washburn1983, Pedroni1976}.\\
Lee-Goldburg decoupling experiments have been conducted at varying offset frequencies in steps of 10 kHz around the centre of gravity of the broad hydrogen spectrum shown in figure 3b. Figure 4 demonstrates the dependence of the linewidth after a Fourier transform of the free induction decay in the rotating frame (FIDRF). As can be seen, a resonance line-width minimum is observed at about foff = 20 kHz.  In these first experiments, the saturating Lee-Goldburg pulses were incremented from initially 1 µs in steps of 500 ns for 8000 increments, resulting in a rotating frame time domain of 4 ms.  \\
\begin{figure}[htb]
  \includegraphics[width=1\columnwidth]{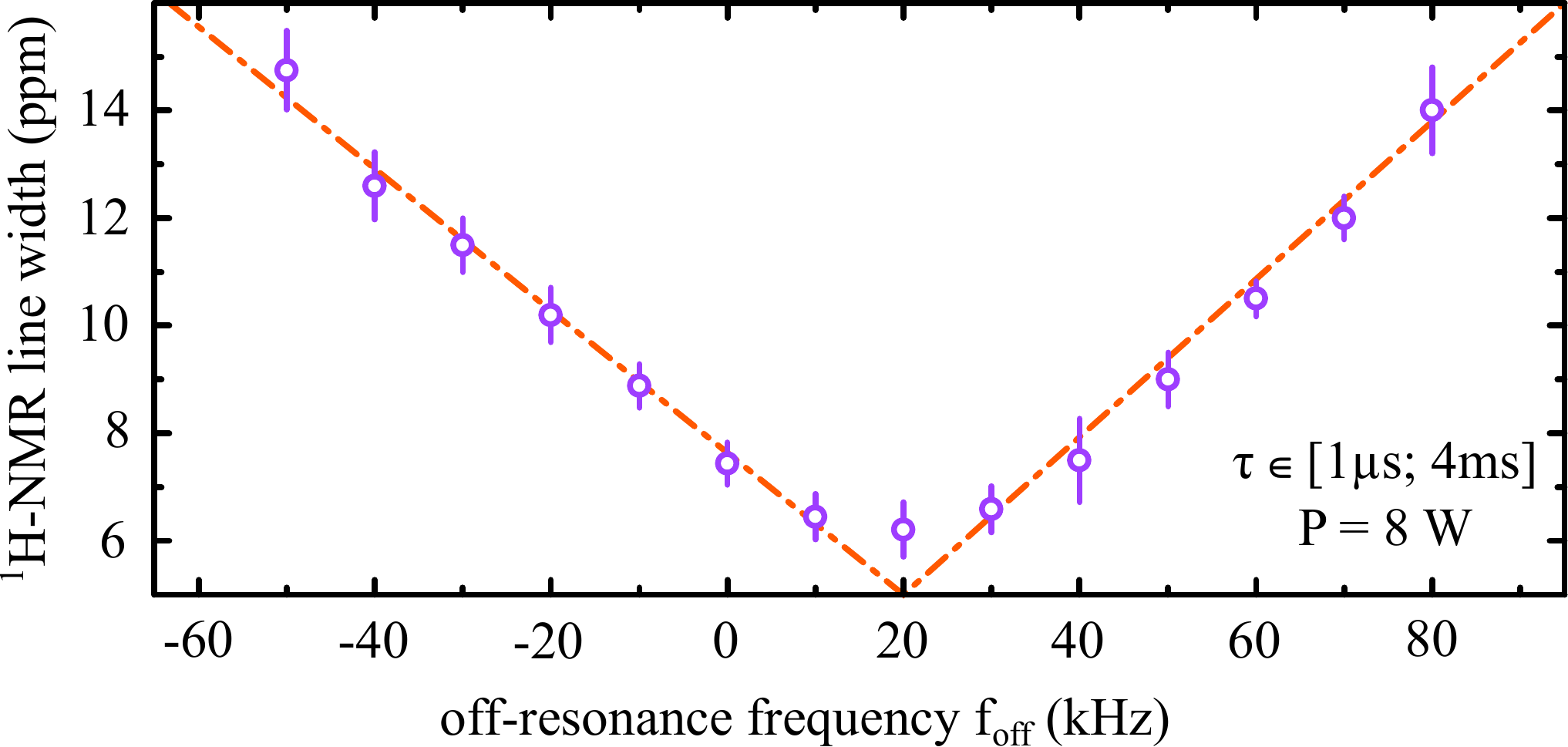}%
 \caption{Dependence of the $^1$H-NMR linewidth after FT of the free induction decay in the rotating frame for different off-resonance frequencies f$_{off}$ at 20 GPa.
 \label{FIG4}}
 \end{figure} 
Figure 3a also shows the resulting FIDRF at 20 GPa. The free induction decay is stretched by a factor proportional to $|\lambda_0(\Theta)|^{- 1}=|\frac{1}{2}(3cos^2 \Theta -1|^{-1}$ where $\Theta$ denotes the angle between the direction of the irradiated LG-field and the external magnetic field B$_0$ acting on the high pressure chamber. Comparison of the $1/e$ decay times, 900 ns for the FID (blue line in figure 3a) and about 2 ms for the FIDRF (orange trace in figure 3a) suggests an elongation of the FID by a factor of about 2000 and thus $|\frac{1}{2}(3cos^2 \Theta -1|^{-1} \approx 4.5 \cdot 10^{-4}$ or $\Theta = 54.717^{\circ}$, close to the magic angle.  \\
As the FIDRF is not fully decayed within the predefined time frame of 4 ms, the minimal line-width of its Fourier transform can be estimated to be $FWHM \approx 1/ (4 ms) = 250~Hz \equiv 5.5~ppm$. Experimentally, we find a line width of about 6.4 ppm. 
By increasing the time scale of the LG-pulse to 16 ms (8000 increments in 2 $\mu$s steps), the FIDRF has been found to be almost completely decayed within the time frame of the LG-pulse, resulting in a spectral resolution of 3.1 ppm (see yellow spectrum in figure 3b).\\
Figure 5 shows a summary of the FWHM line widths of $^1$H-NMR spectra of molecular ortho-hydrogen in a pressure range of 20 to 64 GPa, after standard single pulse excitation and LG-decoupling. In almost all cases, line widths of about 3 to 5 ppm could be realised. \\
All resonance spectra were found to be almost fully represented by a Lorenzian line shape without any detectable signs of chemical shift anisotropy, evidencing the suitability of this approach for a purely dipolar broadened spin-1/2 systems. An interpretation of the chemical shift dependencies at increasing densities will be published elsewhere.\\
The aim of this study was to elucidate the possibility to achieve high spectral resolutions in diamond anvil cell based NMR. It was found that even one of the most basic homo-nuclear decoupling experiments, i.e. Lee-Goldburg irradiation, can be utilised to achieve resolutions as low as 3 ppm. While this value might still be considered too high when compared to magic angle spinning techniques, it represents a major improvement in spectral resolution in diamond anvil cells at extreme conditions.\\
The found resolution of 3 ppm may be improved in future experiments by extending the time domain in the rotating frame, or by achieving a higher SNR per increment by using a higher number of accumulations. 
\begin{figure}[htb]
  \includegraphics[width=.95\columnwidth]{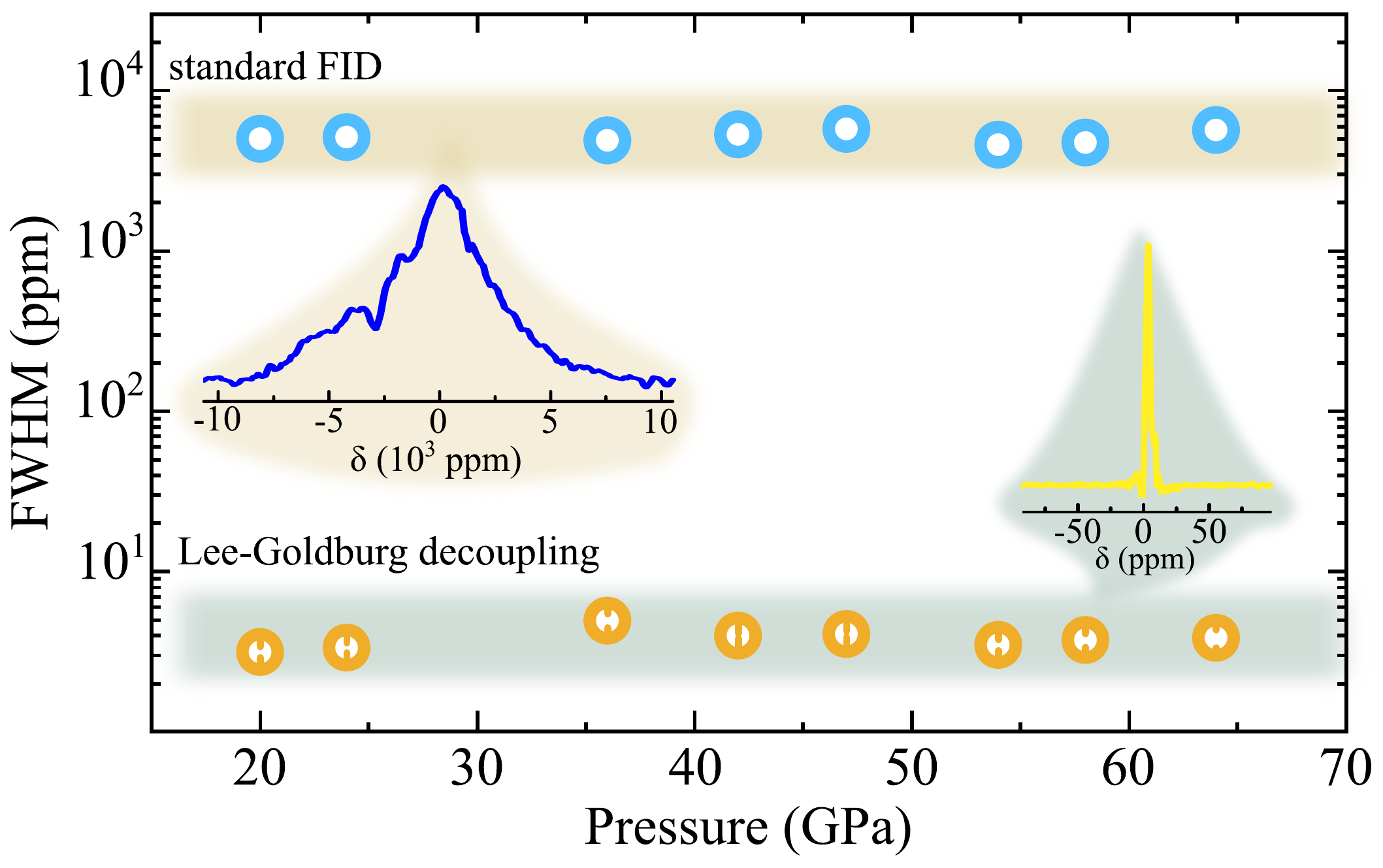}%
 \caption{Comparison of FWHM line widths after standard 90° pulse excitation, and after Lee-Goldburg decoupling in a pressure range from 20 to 64 GPa.
 \label{FIG5}}
 \end{figure}
The applicability of homo-nuclear Lee-Goldburg decoupling is particularly powerful in systems governed by direct dipole-dipole interactions, such as H$_2$, which are almost completely averaged out during the LG-experiments leaving only isotropic chemical shift interactions present. These results open the way for experiments at even higher pressures, and may eventually resolve the predicted structural models of Phases II and III of hydrogen\cite{Freiman2017}.

\section{Acknowledgements}
The authors thank the German Research Foundation (Deutsche Forschungsgemeinschaft, DFG, projects DU 954/11-1, DU 393/13-1, DU 393/9-2, and ME 5206/3-1) and the Federal Ministry of Education and Research, Germany (BMBF, grant no. 05K19WC1) for financial support. 


\section{Supplementary}

\begin{figure*}[htb]
  \includegraphics[width=1.75\columnwidth]{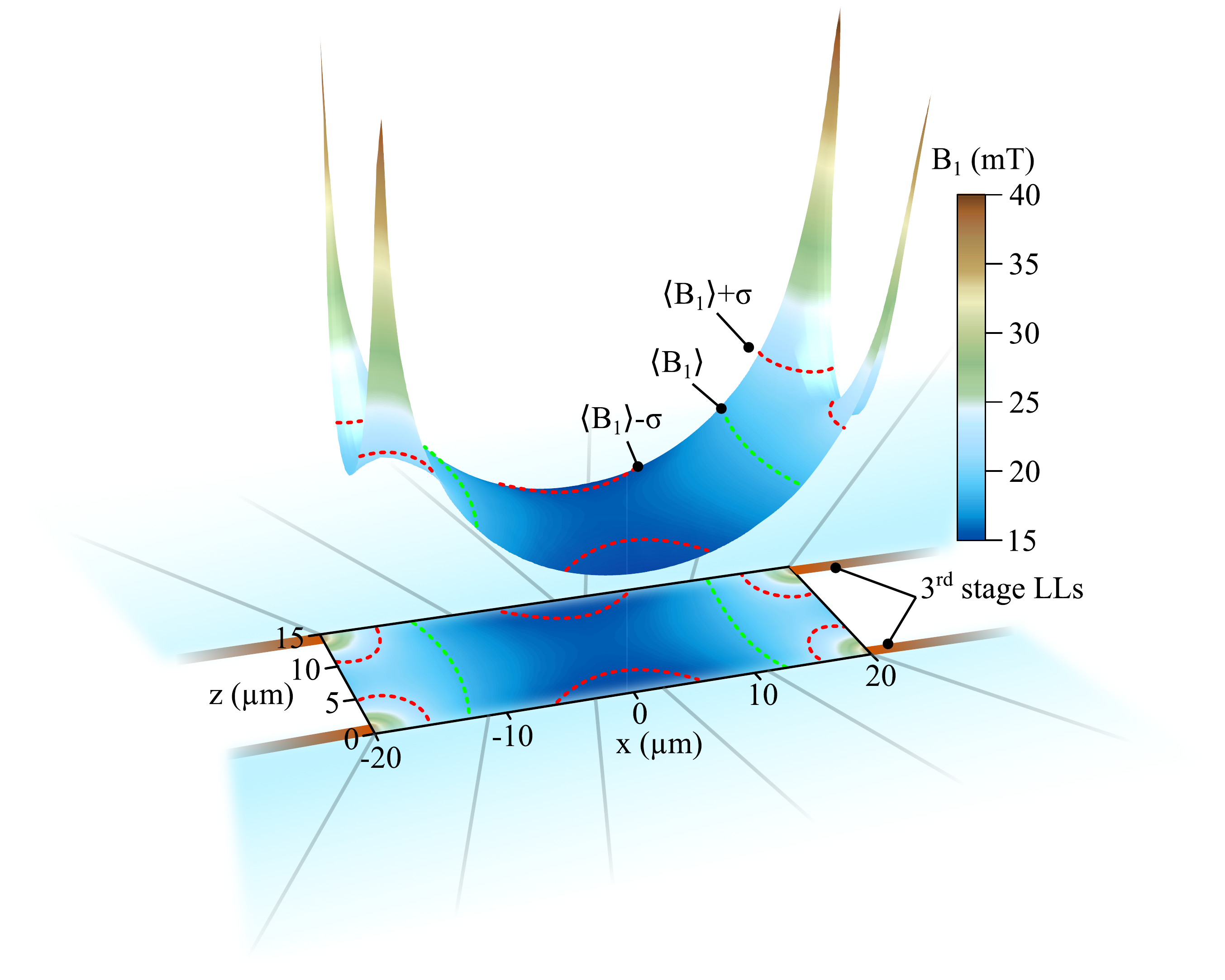}%
 \caption{Finite element simulation of the RF B$_1$ field generated by the 3rd stage Lenz lenses across a 12 pl sample cavity. The green lines denote an average value of about $\braket{B_1} = 17~ mT$. The red lines denote the upper and lower (defined by the standard deviation) boundaries of the effective volume $V_{eff}$.  
 \label{FIGS1}}
 \end{figure*}

\end{document}